\documentclass[runningheads]{llncs}

\usepackage[mobile]{eccv}

\usepackage{eccvabbrv}

\usepackage{graphicx}
\usepackage{booktabs}
\usepackage{multirow}
\usepackage{makecell}
\usepackage{mathabx}
\usepackage[accsupp]{axessibility}  %

\usepackage{wrapfig}

\usepackage[pagebackref,breaklinks,colorlinks,citecolor=eccvblue]{hyperref}

\usepackage{orcidlink}

\definecolor{MyDarkBlue}{rgb}{0,0.08,1}
\definecolor{MyDarkGreen}{rgb}{0.02,0.6,0.02}
\definecolor{MyDarkRed}{rgb}{0.8,0.02,0.02}
\definecolor{MyDarkOrange}{rgb}{0.40,0.2,0.02}
\definecolor{MyPurple}{RGB}{111,0,255}
\definecolor{MyRed}{rgb}{1.0,0.0,0.0}
\definecolor{MyGold}{rgb}{0.75,0.6,0.12}
\definecolor{MyDarkgray}{rgb}{0.66, 0.66, 0.66}
\definecolor{MyDarkCyan}{rgb}{0.05, 0.55, 0.45}
\definecolor{MyBlack}{rgb}{0., 0., 0.}
\definecolor{MyMagenta}{rgb}{1., 0., 1.}
\definecolor{BerkeleyYellow}{RGB}{255,204,41}
\definecolor{BerkeleyLightBlue}{RGB}{94,146,221}
\definecolor{BkDarkBlue}{rgb}{.05,.07,.353}

\newcommand{\ignorethis}[1]{}

\usepackage[title]{appendix}

\begin{document}

\title{MEET: Mixture of Experts Extra Tree-Based sEMG Hand Gesture Identification} 

\titlerunning{MEET}

\author{Naveen Gehlot\inst{1} \quad Ashutosh Jena\inst{1} \quad Rajesh Kumar\inst{1} \\ Mahipal Bukya\inst{2} }

\authorrunning{N. Gehlot et al.}

\institute{Malaviya National Institute of Technology\inst{1}\; \;
Manipal Institute of Technology\inst{2}\; \;}

\maketitle

\begin{abstract}

Artificial intelligence (AI) has made significant advances in recent years and opened up new possibilities in exploring applications in various fields such as biomedical, robotics, education, industry, etc. Among these fields, human hand gesture recognition is a subject of study that has recently emerged as a research interest in robotic hand control using electromyography (EMG). Surface electromyography (sEMG) is a primary technique used in EMG, which is popular due to its non-invasive nature and is used to capture gesture movements using signal acquisition devices placed on the surface of the forearm. Moreover, these signals are pre-processed to extract significant handcrafted features through time and frequency domain analysis. These are helpful and act as input to machine learning (ML) models to identify hand gestures. However, handling multiple classes and biases are major limitations that can affect the performance of an ML model. Therefore, to address this issue, a new mixture of experts extra tree (MEET) model is proposed to identify more accurate and effective hand gesture movements. This model combines 
 individual ML models referred to as experts, each focusing on a minimal class of two. Moreover, a fully trained model known as the gate is employed to weigh the output of individual expert models. This amalgamation of the expert models with the gate model is known as a mixture of experts extra tree (MEET) model. In this study, four subjects with six hand gesture movements have been considered and their identification is evaluated among eleven models, including the MEET classifier. Results elucidate that the MEET classifier performed best among other algorithms and identified hand gesture movement accurately. Leveraging MEET achieves an accuracy of 86.8\%, 89.2\%, 87.9\%, and 78.4\%, for the given four subjects and concludes as the best classifier over other models in comparison.

\end{abstract}

\section{Introduction}
\label{sec:introduction}
The rapid advancement of artificial intelligence technology and the proliferation of intelligent devices across various aspects of life has shown consistent growth. These devices demonstrate capabilities across diverse domains, including industrial production, military operations, healthcare, and service sectors. It is foreseeable that human-computer interactions (HCI) is increasingly permeate future scenarios. In contrast to conventional approaches, HCI grounded in bioelectric signals offers the potential to control external apparatuses, including intelligent machinery, robotic systems, aircraft, and virtual simulations. Hence, exploring the decoding of human bioelectric signals to discern associated behavioral intentions represents a pivotal frontier in the realm of HCI research~\cite{zhang2024hardware}.

The array of human bioelectric signals are electroencephalo-graphy (EEG), electromyography (EMG), and electroneuro-graphy (ENG)~\cite{li2020review}. ENG entails a craniotomy and the implantation of sensors into the brain, a complex procedure that can inflict severe harm on the human body. Moreover, ENG surgery and maintenance require medical expertise, rendering it unsuitable for routine operational usage. EEG signals, though widely studied, are weak and susceptible to environmental interference, thus still primarily in the research phase with limited practical utility. EMG is the method of capturing and displaying signals generated by the body's muscle system. These signals can be acquired using either needle electrodes (iEMG) or surface electrodes (sEMG). The majority opt for non-invasive assessments, such as sEMG. Typically, Ag/AgCl-based electrodes are employed, along with conductive gel, to enhance conductivity on the skin's surface. The muscle's electrical activity is monitored by appropriately positioning the electrode above the epidermis~\cite{wu2019semg}.

sEMG represents the summation of excitation potentials received by neurons, reflecting a person's intention to activate a specific motor muscle. Deciphering sEMG patterns can reveal intentions related to body actions by accurately representing neural information about potential movements. This trait aligns with a key goal of HCI interface design: gathering intentions to influence human actions and creating devices to assist external motions based on those intentions~\cite{shi2020feature}. EMG's versatility and compatibility with human behavior make it ideal for clinical rehabilitation, prosthetic control, and HCI applications such as gesture recognition. Gesture recognition is a significant area of study within biomedical engineering and artificial intelligence. The ability to accurately interpret human hand movements has numerous applications, including prosthetic control and human-computer interfaces.

Human gesture recognition involves essential steps: data acquisition of the hand gesture, preprocessing, and feature extraction, and selecting the most suitable machine learning model to identify the gesture accurately. The initial step towards gesture recognition is data acquisition. His research by Jongman Kim et al.~\cite{kim2021semg} was collected using the Myo armband with 8 channels, performing and analyzing 12 different gestures with an AI model. Jiayuan He et al.~\cite{he2020biometric} utilized sixteen monopolar sEMG electrodes (AM-N00S/E, Ambu, Denmark) placed in pairs around the right forearm to acquire 16 different gestures. Abbas Rahimi et al.~\cite{rahimi2016hyperdimensional} used 4 EMG sensor bases to acquire 5 different hand gesture signals. Rami N. Khushaba et al.~\cite{khushaba2012toward} collected data using EMG channels (Delsys DE 2.x series EMG sensors) and processed it with the Bagnoli Desktop EMG Systems from Delsys Inc. A2-slot adhesive skin interface was applied to each sensor for firm attachment to the skin. This setup was used for 10 different hand gestures and applied a learning model for accurate gesture classification. Xueyan Tang et al.~\cite{tang2012hand} developed a channel acquisition device for multichannel sEMG ring for 11 gesture acquisition. In a study by Suguru Kanoga et al.~\cite{kanoga2020armband}, eight-channel sEMG signals were acquired from the right upper limb of subjects using the Myo gesture control armband (Thalmic Labs, Kitchener, Canada).

The acquired signal from data acquisition fed into the machine learning model requires the extraction of hand-crafted features for more accurate classification. In~\cite{kanoga2020armband}, the study focused on extracting features such as mean absolute value (MAV), zero crossing (ZC), slope sign change (SSC),  waveform length (WL), root mean square (RMS), and autoregressive (AR) coefficients after data acquisition, which were then fed into the classifier to achieve good results. Similarly, in~\cite{khushaba2012toward}, the emphasis was on enhancing the performance of the applied AI model by extracting features such as ZC, WL, SSC, AR, sample skewness (SS), Hjorth time domain parameters (HTD). In the study by~\cite{ariyanto2015finger}, a variety of features were extracted, including Integrated EMG (iEMG), MAV, simple square integral (SSI), variance (VAR), RMS,  difference absolute standard deviation value (DASDV), AR, HTD which were then applied to an AI model to achieve favorable results. Additionally, in the study by~\cite{kim2021semg}, vectors of features such as RMS, MAV, iEMG, ZC, WAMP, VAR, DASDV, LOG, MYOP, AAC, AR, and CC were extracted and fed into the AI model to yield satisfactory outcomes.

Eventually, the extracted features are fed into the AI model, which is a subset of artificial intelligence, typically a machine learning (ML) model. In the study by Ji-Won Lee et al.~\cite{lee2023wearable}, various ML models such as support vector machine (SVM), Ensemble, k nearest neighbour (KNN), naive bayes (NB), and Trees were applied to classify gestures. However, since these models were considered individually, they might be biased and yield better accuracy for some classes than others. Binish Fatimah et al.~\cite{fatimah2021hand} applied SVM, KNN, ensemble bagged trees, and ensemble subspace discriminant to classify gestures, obtaining good results on the acquired data. This study utilized a single model with comparative analysis. Sara Abbaspour et al.~\cite{abbaspour2020evaluation} classified hand gestures using linear discriminant analysis, KNN, decision tree (DT), SVM, multilayer perceptron, etc., and compared them based on accuracy to determine the best classifier. The researcher employed only one classifier for the entire dataset, showing high accuracy for a few classes compared to others. Smita Bhagwat et al.~\cite{bhagwat2020electromyogram} applied KNN, SVM, quadratic discriminant analysis, etc., achieving more accurate classification results. This study uses feature selection and reduction and a mixture network to enhance highly accurate results.

\begin{figure*}[]
\centering
\includegraphics[width=0.95\linewidth, height=0.3\textwidth]{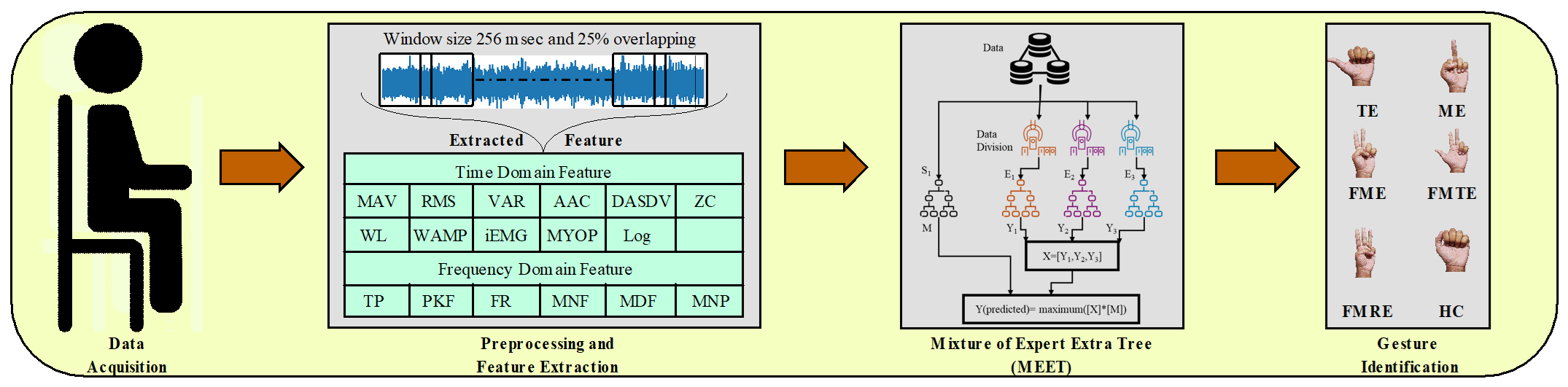}
\caption{Flow architecture for MEET-based gesture classification}
\label{figh1}
\end{figure*}

From the literature, it's evident that the recognition rate of finger movements is greatly influenced by the choice of AI machine learning models, their integration with other networks, and the number of handcrafted features used for the model. Another significant factor affecting data acquisition is the number of channels used. Given these research gaps, this article seeks to make the following main contributions:

\begin{enumerate}
    \item 
    In hand gesture recognition, the proposed Mixture of Experts Extra Trees (MEET) model, which involves experts for binary classification, uses a gated classifier to assign weights to its outputs.

    \item The study employed handcrafted feature extraction, which encompasses seventeen features derived from the acquired signal utilizing a two-channel Biopac MP150 acquisition device.    

    \item 
    A comparative analysis of ten machine learning models with the proposed model MEET has been conducted utilizing graphical representations to analyze performance metrics such as precision, recall, F1-score, and accuracy.
\end{enumerate}

\section{Materials and Methods}
The methodology for the study comprises several small stages, as illustrated in Figure~\ref{figh1}. It involves stage 1 data acquisition, and preprocessing the acquired data before proceeding to stage 2 for handcrafted feature extraction. The extracted handcrafted features from stage 2 are then passed through the machine learning model in stage 3, where the proposed model MEETs classify the different hand gestures. 
\subsection{Data Collection}
In this study, data for hand gesture identification has been collected from consenting participants. Prior to participation, each participant provided written informed consent, confirming their understanding and willingness to participate in the study.

\begin{figure*}[]
\centering
\includegraphics[width=0.95\linewidth, height=0.4\textwidth]{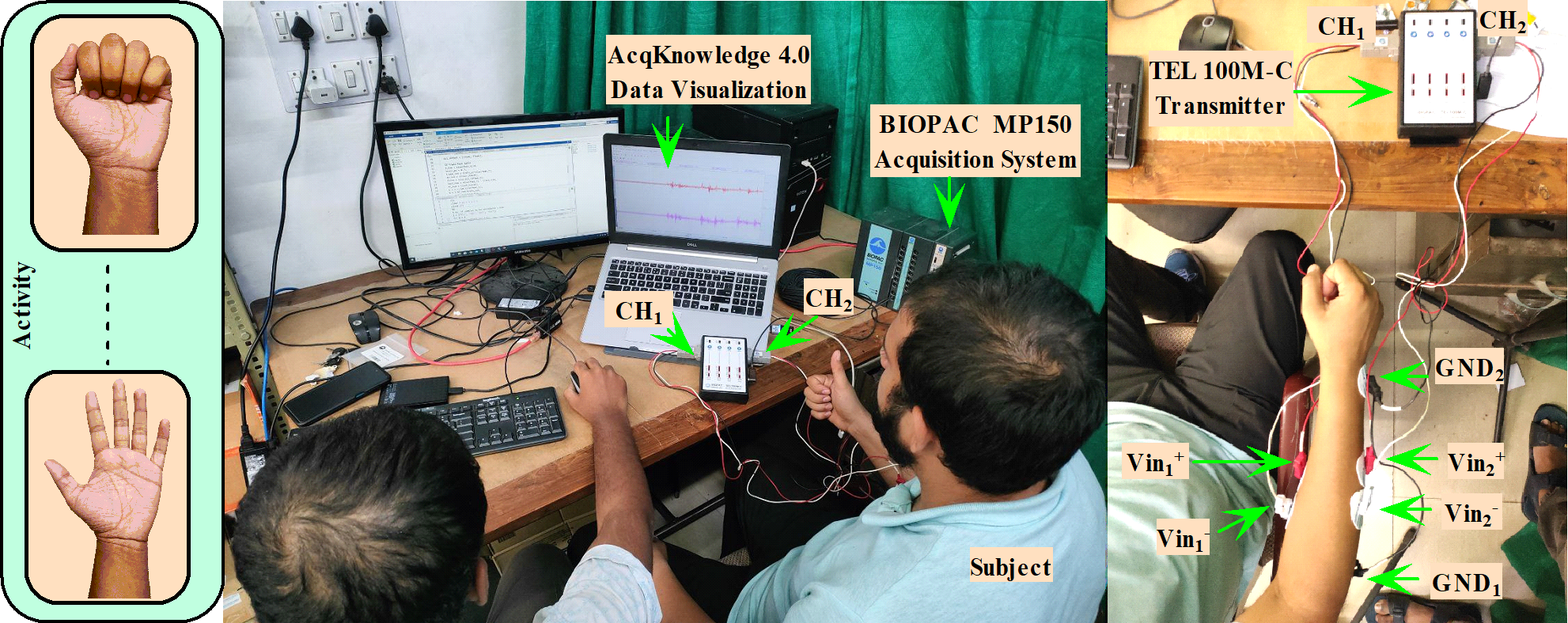}
\caption{Hardware setup used to acquire surface Electromyography (sEMG) data}
\label{figh2}
\end{figure*}

The EMG signals has been acquired using AcqKnowledge software version 4.4, which is paired with the BIOPAC MP150 system known for its efficient EMG data collection and signal processing capabilities. Figure~\ref{figh2} illustrates the setup utilized throughout the data acquisition process. It displays the hardware arrangement for data gathering, which involves placing disposable Ag/Ag-Cl surface electrodes on the participant's forearm muscle to capture the sEMG signal. Signal evaluation primarily focuses on the extensor digitorum and flexor pollicis longus muscles~\cite{shi2018bionic, hamrick1998emg}. The sEMG signals were recorded using a shielded cable electrode connected to a TEL-100MC filter amplifier with a gain of 1000. This configuration facilitated signal amplification and noise reduction. The data has been captured at a 2000 Hz sample rate to ensure high-frequency resolution and an accurate depiction of muscle activation during hand gestures.

Four healthy adult male and female subjects participated in this study. The participants were instructed to perform activities in a predetermined order, with six actions being examined. Each of these exercises has been consisted of three steps, with participants advised to maintain a state of meditation and relaxation throughout the working phase.

\textbf{Stage I:} At this stage, participants are briefed on the exercise two minutes before commencing the assignment.

\textbf{Stage II:} At this stage, participants are instructed to perform the task continuously for 40 seconds.

\textbf{Stage III:} Participants are given time to recuperate after each activity. Muscle signals may decrease as the muscles become fatigued.

\subsection{Preprocessing and Feature Extraction}
During the data-collecting phase, sEMG signals are sensitive to undesirable external effects, often known as noise. Noise can take many forms, including power line interference, electromagnetic noise, and interference from electronic devices like phones and broadband equipment~\cite{chowdhury2013surface}. To assure data quality, filtering techniques are used to decrease noise. A notch filter is used to remove the power line frequency, which is commonly 50Hz. In addition, a Bandpass filter with a cutoff frequency between 10Hz and 500Hz is used. These filtering techniques use mathematical tools to decrease the influence of noise and make the original sEMG signals more suitable for further investigation.

This filtered data is then utilized for feature extraction, which is an important component in the performance of machine learning models. Feature extraction includes segmenting the sEMG signal into 256-millisecond windows with a 25\% overlap between successive windows~\cite{r_15}. This windowing approach achieves two goals: it reduces data duplication and tackles the difficulty of maintaining immediate signal information. 
Six frequency domain features and eleven time domain features are extracted within these segmented windows, respectively. The utilized handcrafted features include MAV, VAR, DASDV, WL, iEMG, Log, RMS, AAC, ZC, WAMP, and MYOP, which are time domain features, and total power (TP), frequency ratio (FR), median frequency (MDF), peak frequency (PKF), mean frequency (MNF), and mean power (MNP) as frequency domain features~\cite{gehlot2023semg}. These features are crucial for advancing research in hand gesture detection.

\subsection{ML model}
This study employs ten machine learning models with the proposed model MEET to classify different hand gestures. Performance metrics for the utilized models are examined, and the best classifier is chosen to aid enhance gesture recognition. Machine learning methods for recognizing hand gestures include Decision Tree (DT), Random Forest (RF), Gradient Boosting (GB), Adaboost (ADB), Support Vector Machine (SVM), Bagging (BAG), Logistic Regression (LR), Naive Bayes (NB), k-Nearest Neighbor (KNN), Extra Tree (ET), and Mixture of Experts Extra Tree (MEET). The short description of the examined machine learning models is as follows:

\subsubsection{Decision Tree (DT)}
To categorize data, decision trees employ a basic, straightforward if-then rule paradigm. This involves constructing a hierarchical tree structure starting from the root node, which is then divided into additional nodes, connected by branches. These branches are further subdivided until they terminate at leaf nodes, where final classification decisions are made. By following this methodology, the Decision Tree model effectively captures and visualizes underlying patterns and relationships within the data, enhancing the comprehensibility and accuracy of classification outcomes. The systematic progression from root to leaf nodes illustrates the categorization process in a clear and structured manner, facilitating its application in data analysis tasks.

The data input comprises a set of input instances, denoted as S, each associated with a probability distribution $p_{d1}$, $p_{d2}$, ..., $p_{dM}$, and labeled with classes 1, 2, 3, ..., M. The entropy of a model is defined as:
\begin{equation}\label{e4}
    Entropy(E) = \sum_{i=1}^{M}{-p_{di}log_2p_{di}}
\end{equation}

Node splitting is determined by maximizing information gain, given by:
\begin{equation}\label{e5}
    IG(E) = Entropy(E)-\sum_{i=1}^{M}\frac{\left | E_i \right |}{\left | E \right |}Entropy(E_i)
\end{equation}

Information gain is calculated for each feature of the selected data input to identify the feature with the highest information gain. Attributes are then sorted based on their information gain values, and a decision tree (DT) is constructed. At each node, the attribute with the highest gain among the features not yet considered in the current route from the root is chosen~\cite{quinlan1996improved}.

\subsubsection{Random Forest (RF)}
Random Forest is a form of ensemble learning suitable for both classification and regression tasks. The strength of an ensemble is determined by the quantity of decision trees within the forest. Unlike individual decision trees, which can tend to overfit the data and yield less accurate results, Random Forest addresses this issue effectively. Random Forest mitigates overfitting by aggregating predictions from multiple trees while preserving model fidelity~\cite{breiman2001random, jahangiri2015investigating}.

\subsubsection{Gradient Boosting (GB)}
Gradient Boosting operates under the assumption of employing weak learners. This algorithm progressively combines weak learners iteratively to enhance them into strong learners. The objective of gradient boosting is to develop a base learner that closely approximates the negative gradient of the loss function~\cite{friedman2002stochastic,friedman2001greedy}.

Let us suppose that there are $m$ samples.
${(x_1,y_1). . .(x_m,y_m)}\;\;\;\; (\therefore x_i\epsilon x\subseteq R_n,\; and \; y_i\epsilon y\subseteq R)$.
The function for additive approximation is outlined as follows:
\begin{equation}
    f_k(x):= \sum_{k=1}^{k}T(x;\theta_k)
\end{equation}
where, $T(x;\theta_k)$ is the new decision tree and $\theta_k$ is the minimization of the risk parameter of the new decision tree which is explained:
\begin{equation}
\hat{\theta_k}=argmin\sum_{i=1}^{m}L(y_i,f_{k-1}(x)+T(x;\theta _k))
\end{equation}
Within this algorithm, the ultimate forward mode is articulated as:

\begin{equation}
    f_k(a)=f_{k-1}(x)+T(x;\theta _k)
\end{equation}
Ultimately, the value of the loss function's negative gradient is utilized to determine the residual:

\begin{equation}
    R_{ki}=-\left [ \displaystyle\frac{\partial L(y_i,f(x_i))}{\partial f(x_i)}  \right ]
\end{equation}

\subsubsection{Adaboost (ADB)}
Adaboost is a type of ensemble learning algorithm wherein multiple classifiers are combined to mitigate the misclassification rate (error rate) of a weak classifier. The primary concept involves amalgamating diverse predictions to construct a robust classifier, leveraging the predictions of numerous weak classifiers on the initial dataset. Bootstrap aggregation (Bagging) and boosting represent pivotal strategies in ensemble classifiers~\cite{schapire1998improved}.

At iteration $p$, it provides the classifier $h_{p}$ and computes the coefficient $\alpha_{p}$. The output of $h_{p}$ is binary, belonging to $\{-1,1\}$, encompassing the classes. When $h_{p}=1$, the weak learner casts a vote for the class; otherwise, it votes against the classes. Following the $p^{th}$ iteration, the output discrimination function is delineated as per~\cite{freund1997decision}.

\begin{equation}
    g(a)=\sum_{p=1}^{P}\alpha_{p}h_{p}(a)
\end{equation}
Whereas,
\begin{equation}
    \alpha_{p} = \frac{1}{2}\log_{e}(\frac{1-\varepsilon }{\varepsilon })
\end{equation}
Here $\alpha_{p}$ is the performance parameter and $\varepsilon$ is the total error. 
For a single label take the class with ``major vote" is selected in further steps.
\begin{equation}
    f(a)=argmax_{l} \; \; g(a)
\end{equation}

\subsubsection{Support Vector Machine (SVM)}
Support Vector Machine distinguishes classes by minimizing classification error while maximizing geometric margin. Support Vector Machine is also referred to as the maximum margin classifier based on structural risk minimization. In this process, the input data dimension space utilizes hyperplanes to discern data separation. To partition the data, two parallel hyperplanes are drawn on each side of the data discrimination. The hyperplane employed for data separation represents the maximum distance between these parallel hyperplanes~\cite{cristianini2000introduction}. The hyperplane utilized to segregate data is outlined as follows:

\begin{equation}\label{ew}
    w.a + b = 0
\end{equation}

\subsubsection{Bagging (BAG)}
Bagging refers to bootstrap aggregation, which is an ensemble-based model. Bagging is the process of creating a succession of independent observations of equal size and distributing them as actual data. It primarily requires two phases in the model:

\begin{enumerate}
    \item Data generation involves transferring each sample data to the base model.
    \item The strategy combines the predictions of the number of predictors.
\end{enumerate}
This classifier is the result of merging many weak classifiers. An accurate prediction model is created by using a large number of classifiers. Multiple decision trees are generated and then combined to form the classifier. The classifier selects a representative sample from the available data and calculates the probability based on it. The probability is estimated by adding and averaging all the unique trees~\cite{breiman1996bagging}.

\subsubsection{Logistic Regression (LR)}
Logistic regression is an advanced linear regression technique that can classify both linear and nonlinear data. This paradigm aids in categorizing data into binary categories. In this case, determining the coefficient value for classification is similar to linear regression. By using the coefficient, it converts the input characteristics into a value between 0 and 1. It is a machine learning approach that involves multiplying the input data by the associated weight. It is sometimes referred to as a discriminatory classifier~\cite{indra2016using}.

\subsubsection{Naive Bayes (NB)}
The Naive Bayes classifier, a supervised machine learning algorithm, operates on the principles of Bayes' rule, a probabilistic approach to learning that leverages existing knowledge. In this context, the maximum a posteriori rule is employed to classify the data. Given data $(B)$ and class $(M)$, the Bayes classifier estimates the parameters $P(M)$ and $P(B|M)$. This rule can be formulated as follows: 
\begin{equation}
    P(M|B)= \frac{P(B|M)P(M)}{P(B)}
\end{equation}
After training the model, the primary goal is to predict the proper class for fresh data. The highest possible posterior class for the given case is:

\begin{equation}
    M_{map} = arg\underset{r\in M}{max}P(B|r)P(r)
\end{equation}
Where $M$ represents the set of all class targets and the features from the input data are denoted as ($b_{1}...b_{n}$), the maximum a posteriori class can be expressed as:

\begin{equation}
    M_{map} = arg\underset{r\in M}{max}P(b_{1}...b_{n}|r)P(r)
\end{equation}
To estimate $P(M)$, the relative frequency of each target class in the training data is calculated.

Estimating $P(b_{1}..b_{n}|M)P(M)$ can be challenging due to the lack of sufficient instances in the training set for every attribute combination, leading to a sparse data problem. To address this issue, the independent assumption is employed, which assumes that attributes are conditionally independent given the target class value~\cite{indra2016using}. Mathematically, this assumption is stated as follows:

\begin{equation}
    P(b_{1}...b_{n}|M)P(M)=\prod_{i}P(b_i|M)
\end{equation}
The subsequent output of the classifier:
\begin{equation}
    M_{NB}=arg\underset{r\in M}{max}P(M)\prod_{i}P(b_i|M)
\end{equation}

\subsubsection{k-Nearest Neighbour (KNN)}
The k-Nearest Neighbour is a popular supervised learning approach that is straightforward to apply. Using this; data may be classed by identifying the surrounding data points. To identify a new data point's class, it calculates the distance between it and all of its nearest neighbors. For example, select the neighbor in the training dataset that is closest to the new data point. Neighbours can use functions such as hamming, Euclidean (EU), Manhattan, Minkowski, and others to calculate this information depending on their distance.

The EU's mathematical representation is as follows:

\begin{equation}
    EU = \sqrt{\sum_{k=1}^{M}({a_k}-{b_k})}
\end{equation}
where, $M$ is the number of attributes or dimensions. ${a_k}$, and ${b_k}$ are the ${k^{th}}$ element object of $a$ and $b$ respectively. The identification of patterns and the mining of data are two of its primary applications.~\cite{dudani1976distance}.

\subsubsection{Extra Tree (ET)}
Extra Tree refers to highly randomized decision trees. This ensemble has the same number of decision trees as the random forest, but it differs in how randomization is implemented during training. The key differences between this method and a random forest are branch training and splitting. The ET algorithm employs the traditional top-down strategy to create an ensemble of the unpruned decision or regression trees. Its two main differences from earlier tree-based ensemble techniques are that it divides nodes at random and creates trees from the complete learning sample (rather than a bootstrap replica).

The method primarily relies on two parameters: $P$, the number of attributes randomly selected at each node, and $n_{min}$, the minimum sample size required for splitting a node. In this context, an attribute denotes a specific input variable used in an ensemble tree (ET). The ensemble parameters, comprising $M$, $P$, and $n_{min}$, exert distinct influences: $P$ governs the robustness of the attribute selection process, $n_{min}$ affects the robustness of averaging output noise, and $M$ determines the robustness of variance reduction in the ensemble model aggregation. These parameters can be adjusted either automatically or manually to suit the specific requirements of the problem. Notably, information gain is normalized in the measured score for classification. This measure can be computed for a sample $A$ and a split $a$ as follows:

\begin{equation}
S_c(a,A)=\frac{2I_c^a(A)}{H_a(A)+H_c(a)}    
\end{equation}
Where $H_c(a)$ is the log entropy for the classification, $H_a(A)$ is the split entropy, and $I_c^a(A)$ is the mutual information of the split outcome and the classification. The algorithm's remarkable efficiency directly results from these properties~\cite{geurts2006extremely,r_13}.

\subsubsection{Proposed Model: Mixture of Experts Extra Tree (MEET)}
Due to a number of limitations, mixture classifiers are used instead of a single one. One main reason is their potential to improve precision. Furthermore, the classifier's mixture structure allows for independent training of sibling modules, making parallel training possible, reducing total training time, and improving accuracy. Single classifiers, on the other hand, contain fewer parameters and are less prone to overfitting, allowing for high accuracy with minimum hyperparameter modification.

\begin{wrapfigure}{r}{6cm}

\includegraphics[width=6cm]{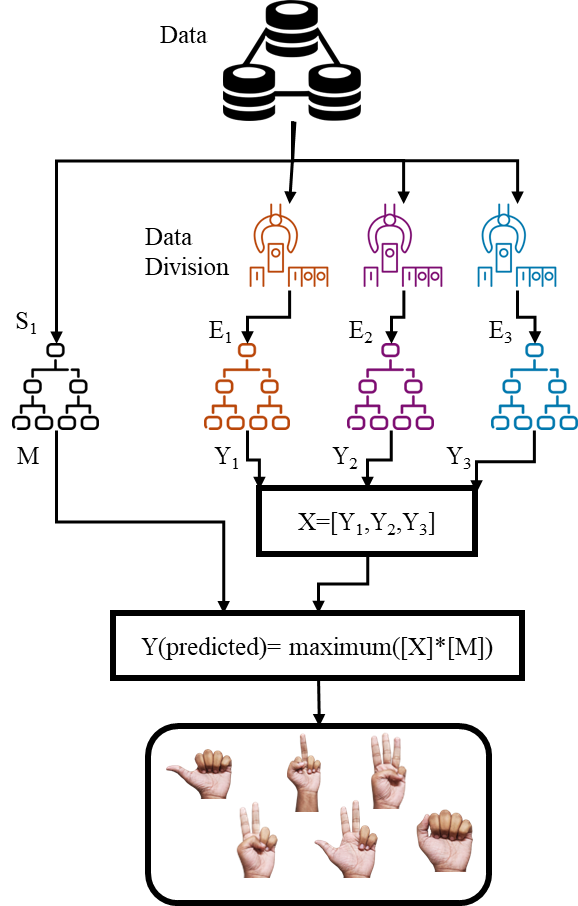}
\caption{MEET: Mixture of Experts Extra Tree Architecture}
\label{figh3}
\end{wrapfigure}

Hence, in this research, the proposed Mixture of Experts Extra Trees (MEET) is structured as depicted in Figure~\ref{figh3}. The figure illustrates that the highly accurate classifier is arranged to classify with minimal classification classes and get a more accurate output. Therefore, the minimum number of classifiers required is determined based on the following equation.
\begin{equation}
    MEET_{classifier}= \frac{N}{2}+1\label{e1}
\end{equation}
Where N represents the number of classes in the data sample.

The $MEET_{classifier}$ is selected as an integer rather than a fraction. In the equation, the additional 1 signifies the classifier that assigns weighted probabilities to all the base classifiers and chooses the most accurate output based on the probability multiplier.

The operational framework of the MEET architecture operates as illustrated in Figure~\ref{figh3}. Initially, the number of the network's classifiers is determined in accordance with equation 18. In this particular scenario involving 6 distinct classes, the requisite classifier comprises 4 classifiers. This classifier comprises 3 expert classifiers, each adept at specific subsets of classes, and 1 gate classifier responsible for adjudicating the weightings attributed to the outputs of the expert classifiers.

The data undergoes an initial segmentation phase, meticulously partitioned by class assignment to each expert classifier. This segmentation ensures that each expert classifier trains on data pertinent to its designated classes, thus circumventing any overlap or confusion during the training process. For instance, Expert E1 is entrusted with the training data corresponding to classes 1 and 2, Expert E2 handles classes 3 and 4, and Expert E3 manages classes 5 and 6. Simultaneously, gate classifier S1 is trained using the entirety of the class data, facilitating its role in orchestrating the final output.

Ultimately, the output of the MEET architecture is a composite of the individual expert classifier outputs—Y1 for E1, Y2 for E2, and Y3 for E3—alongside the weighted output, denoted as M, generated by gate classifier S1. This holistic output is mathematically formalized according to a specific equation, which encapsulates the integration of the expert and gate classifier outputs to yield the MEET architecture's final prediction or decision output shown in equation~\ref{e19}

\begin{equation}
    Y(predicted)=max([[X]*[M]])\label{e19}
\end{equation}

\section{Results and Discussion}

To validate the proposed model, it was applied to datasets from four individuals, each containing 480,000 raw samples, resulting in a total of 1,920,000 samples collected. These datasets are collected by performing six gestures TE, ME, FME, FMTE, FMRE, and HC. These datasets were then applied to the proposed model, MEET, and tested against existing models, including DT, RF, GB, ADB, SVM, BAG, LR, NB, KNN, and ET. All these models were trained on existing data using a 70\% training and 30\% testing split.
\begin{wraptable}{r}{6cm}
\centering
\renewcommand{\arraystretch}{1.5}
\caption{Performance metrics for Machine Learning classifiers}\label{t1}
\begin{tabular}{c|c}\hline \hline
\begin{tabular}[c]{@{}l@{}}Performance Parameter\end{tabular} & Formula \\ \hline \hline
$A_{cc}$ & $\frac{Tp+Tn}{Tp+Fp+Tn+Fn}$ \\
$P_{re}$ & $\frac{Tp}{Tp+Fp}$ \\
$R_{ec}$ & $\frac{Tp}{Tp+Fn}$ \\
$F_s$ & $\frac{2*Pre*Rc}{Pre+Rc}$\\\hline \hline
\end{tabular}
\flushleft \footnotesize{
Here, $A_{cc}$ represents the Accuracy parameter, $P_{re}$ denotes the Precision parameter, $R_{ec}$ stands for the Recall parameter, and $F_s$ represents the F1-score parameter of the performance. Also, $Tp$ represents the true positive, $Fp$ indicates the False positive, $Tn$ represents the true negative, and $Fn$ represents the false negative~\cite{vijayvargiya2021hybrid}.}
\end{wraptable}
The model testing was evaluated based on performance metrics, with each classifier's performance assessed using the parameters listed in Table~\ref{t1}. These performance metrics were calculated using a method known as the confusion matrix. This matrix summarizes the correctly and incorrectly predicted values of each class, allowing the extraction of performance parameters for each class.

Figure~\ref{c1} depicts confusion matrices, with the horizontal axis representing the actual label and the vertical axis representing the predicted label. The confusion matrix provides information about the actual percentage of prediction and classification error when the input is fed into the algorithm. The diagonal values indicate the accuracy of the forecast for each gesture. 
When examining the full matrix, the distribution of diagonal values from left to right downward serves as a measure of the forecast's accuracy.
\begin{figure*}[]
	\centering
	\begin{subfigure}{0.22\linewidth} 
		\includegraphics[width=0.99\textwidth, height=0.9\linewidth]{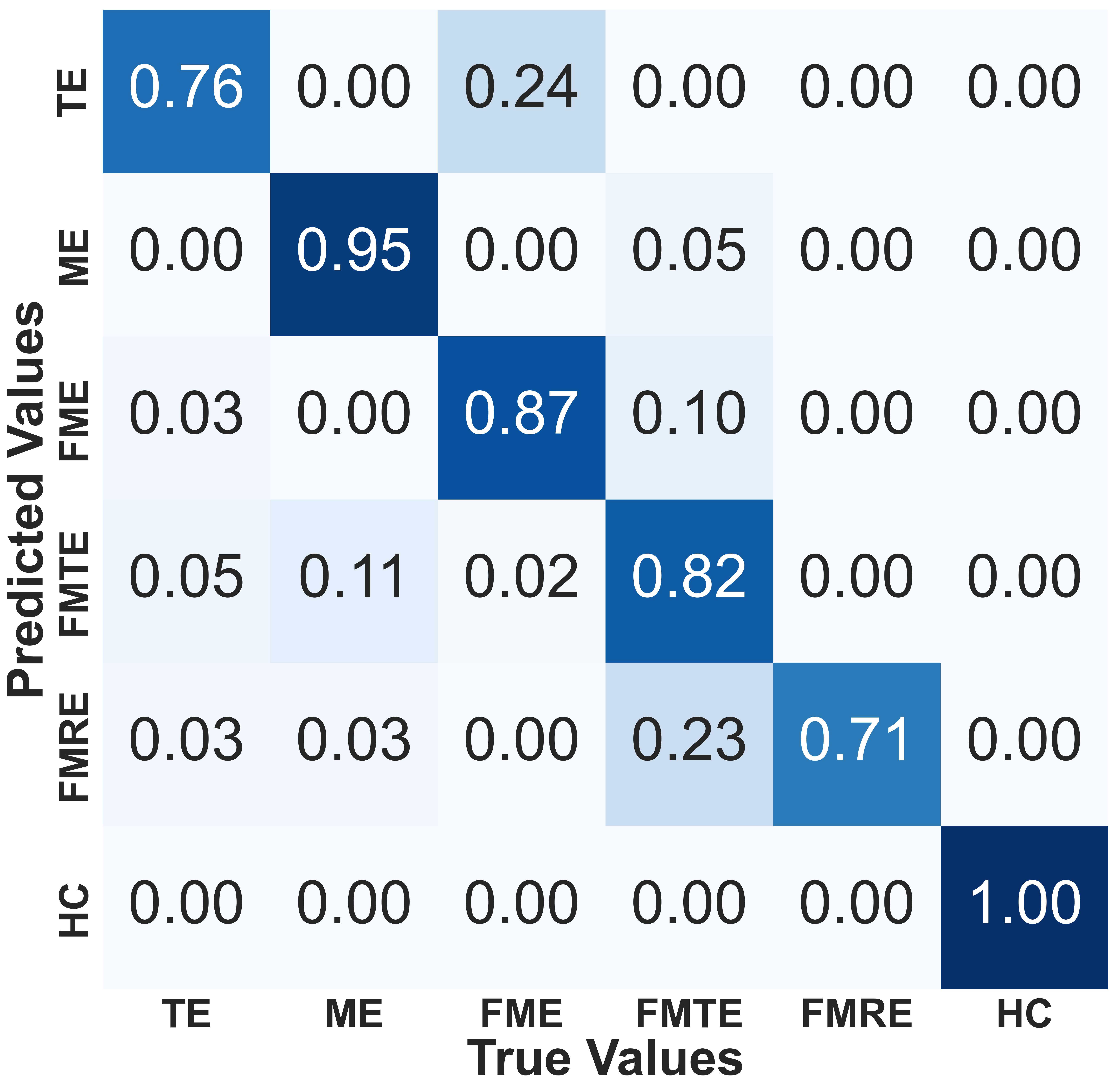}
		\caption{M1}\label{a1}
	\end{subfigure}
	\begin{subfigure}{0.22\linewidth} 
		\includegraphics[width=0.99\textwidth, height=0.9\linewidth]{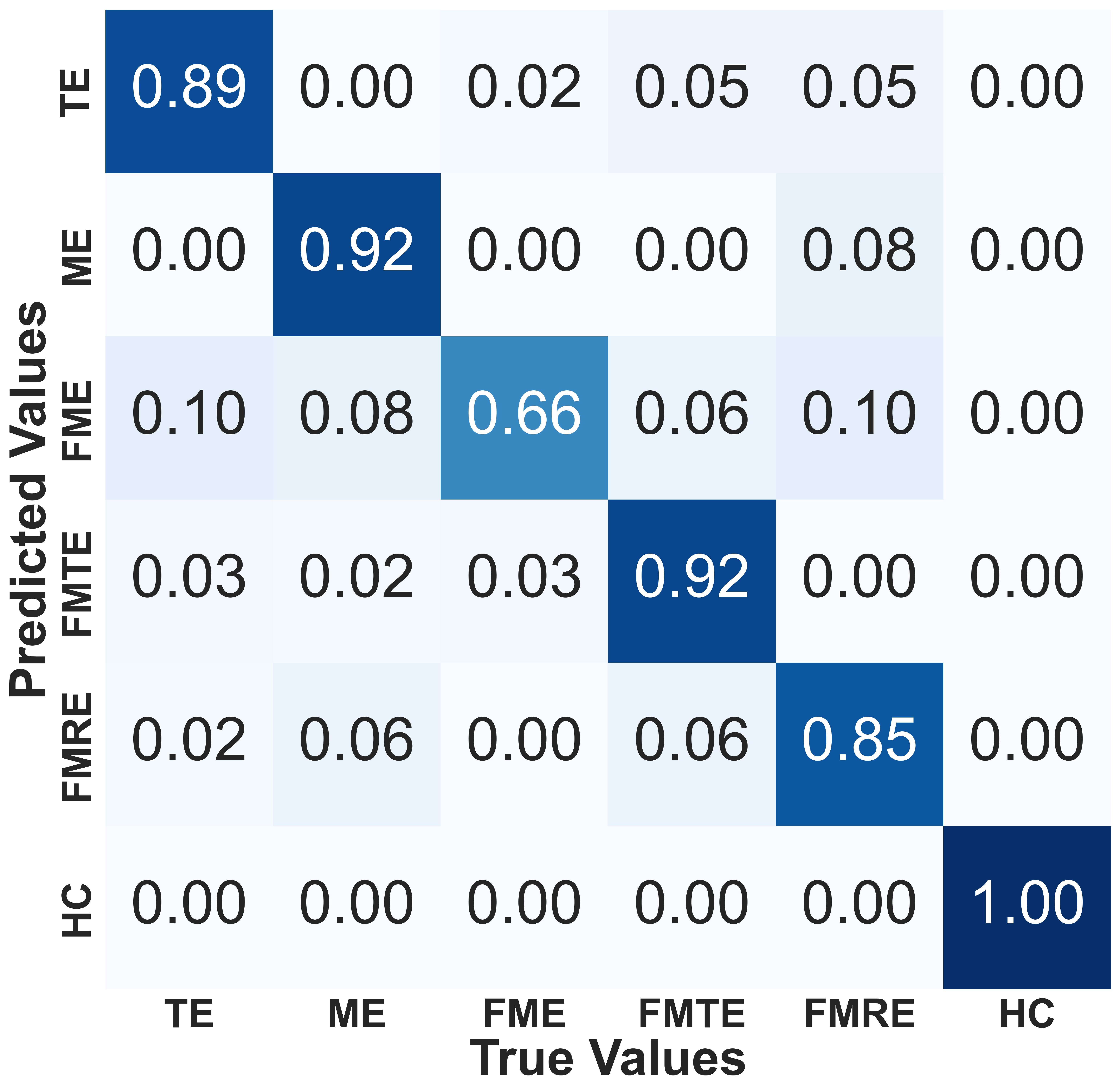}
		\caption{M2}\label{b1}
	\end{subfigure}
	\begin{subfigure}{0.22\linewidth} 
		\includegraphics[width=0.99\textwidth, height=0.9\linewidth]{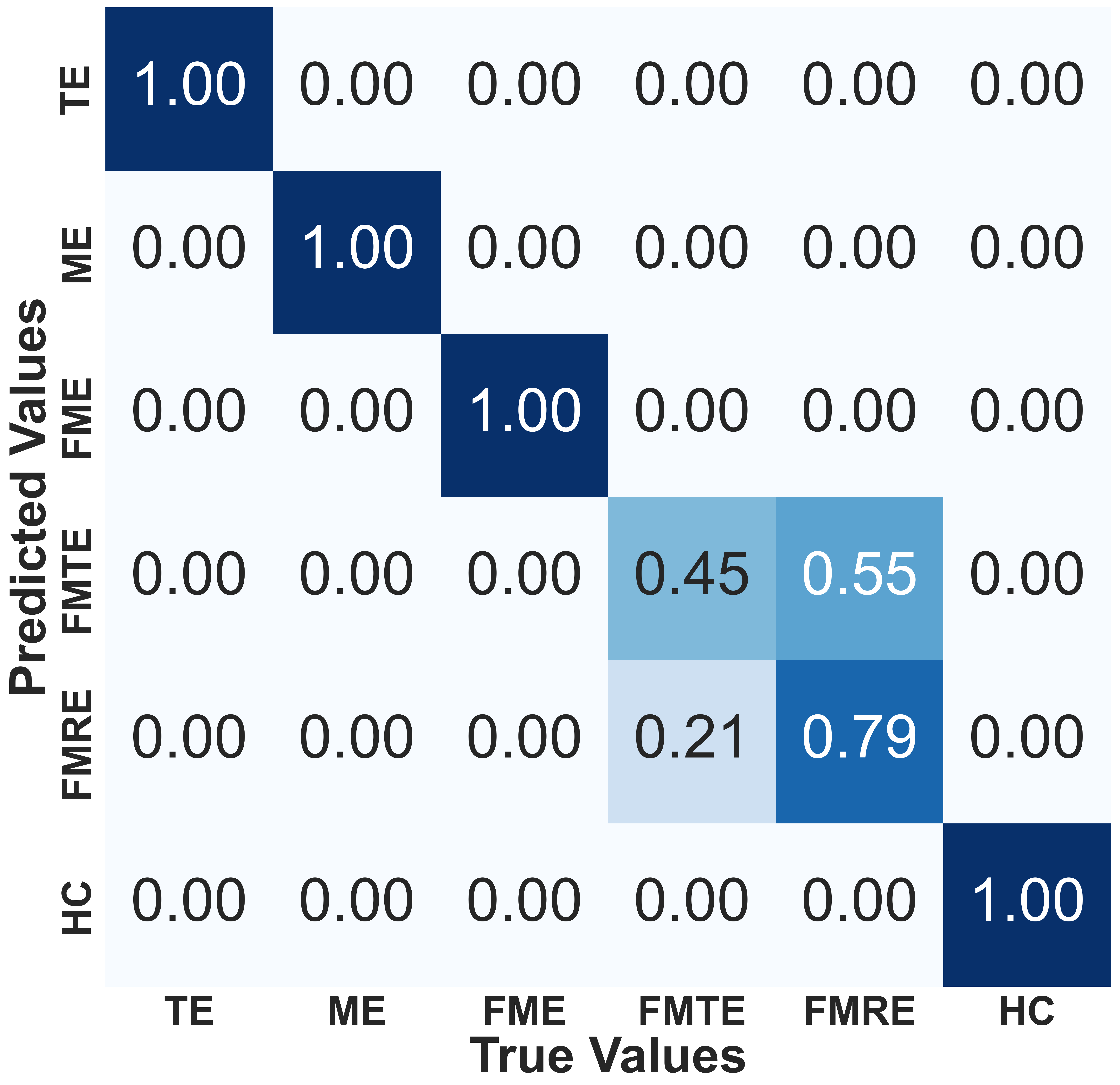}
		\caption{F1}\label{c1}
	\end{subfigure}	
	\begin{subfigure}{0.22\linewidth}
		\includegraphics[width=0.99\textwidth, height=0.9\linewidth]{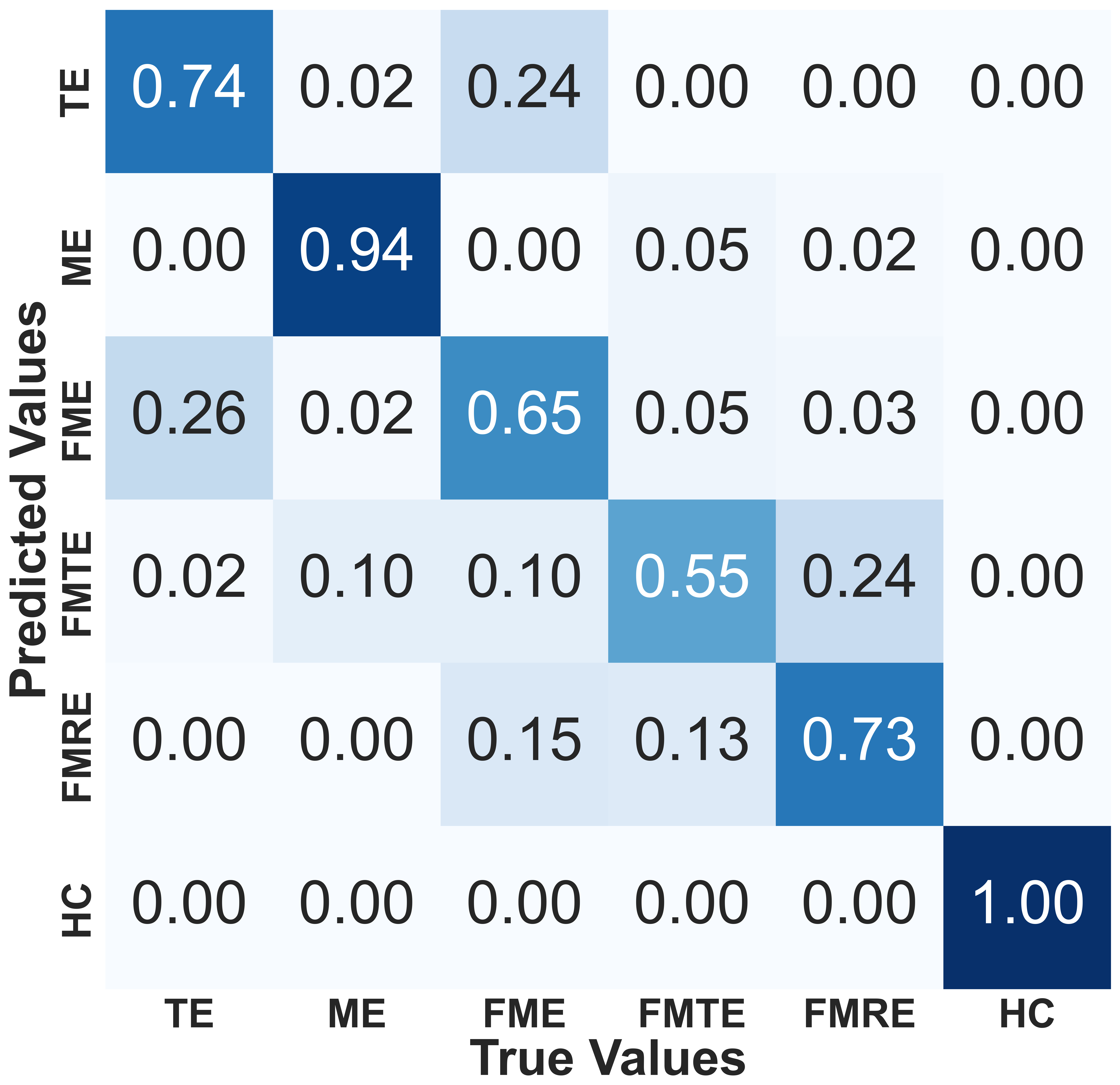}
		\caption{F2}\label{d1}
	\end{subfigure}	
	\begin{subfigure}{0.22\linewidth}
		\includegraphics[width=0.99\textwidth, height=0.9\linewidth]{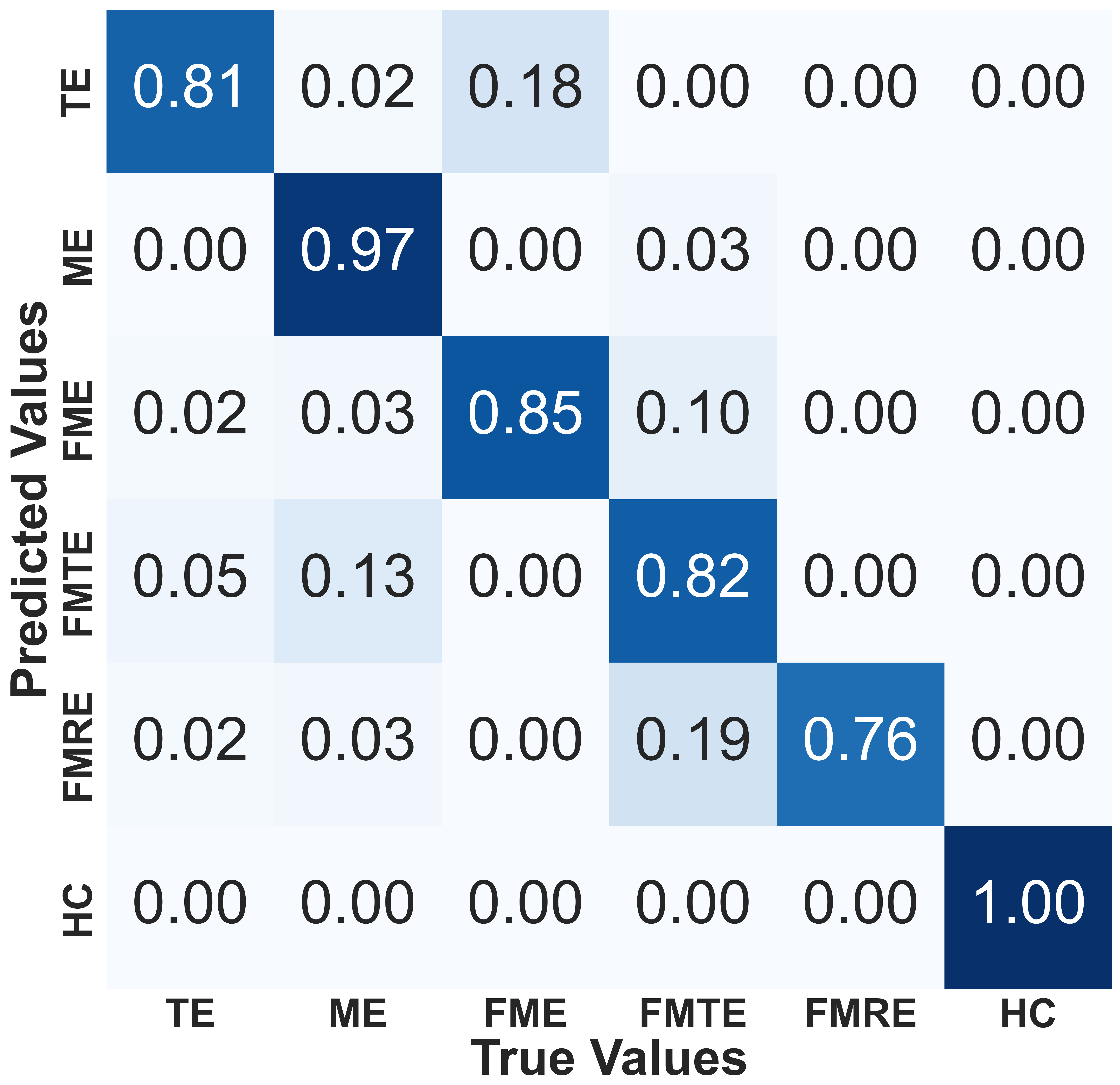}
		\caption{M1}\label{e1}
	\end{subfigure}
	\begin{subfigure}{0.22\linewidth} 
		\includegraphics[width=0.99\textwidth, height=0.9\linewidth]{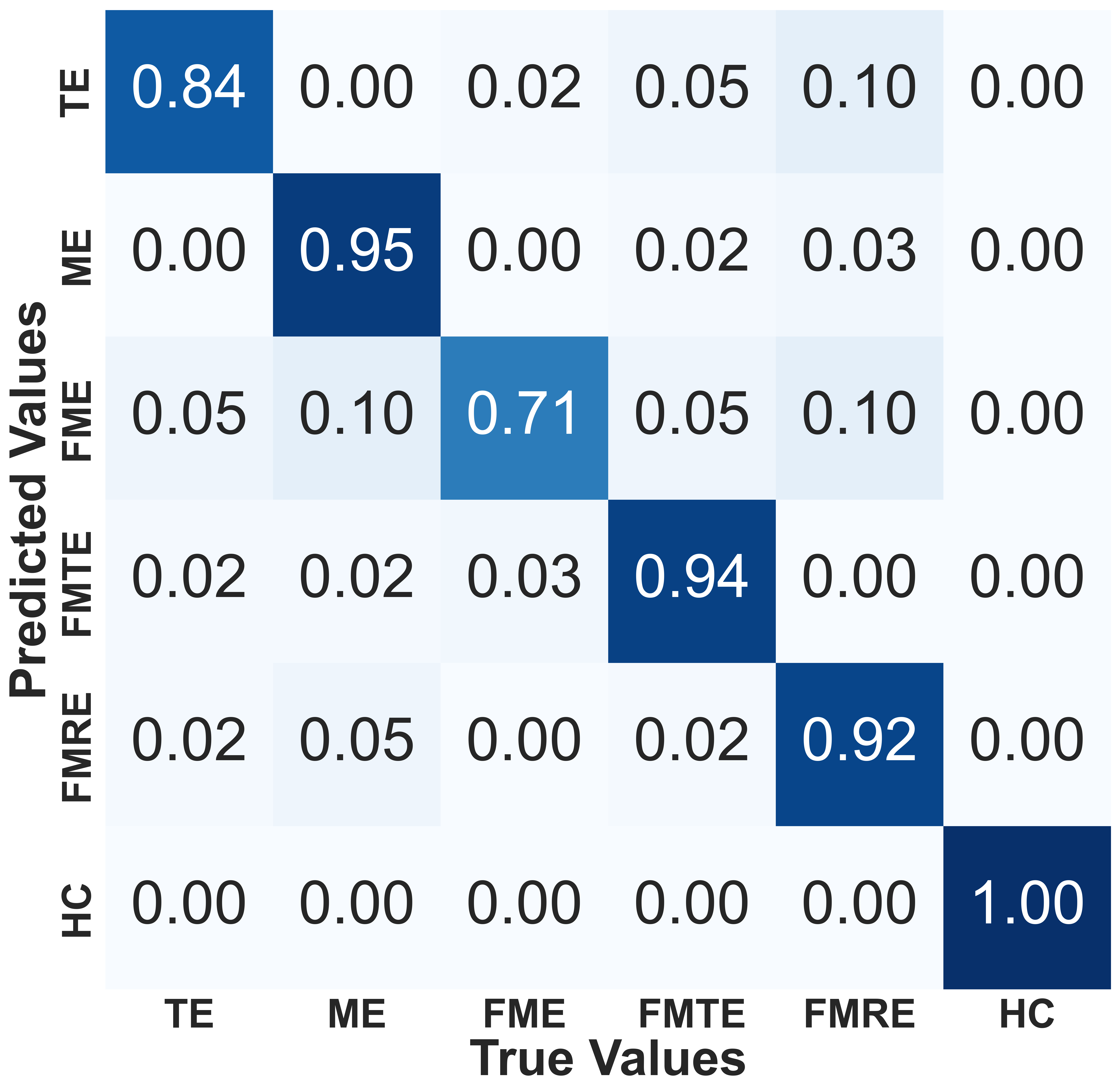}
		\caption{M2}\label{f1}
	\end{subfigure}
	\begin{subfigure}{0.22\linewidth}
		\includegraphics[width=0.99\textwidth, height=0.9\linewidth]{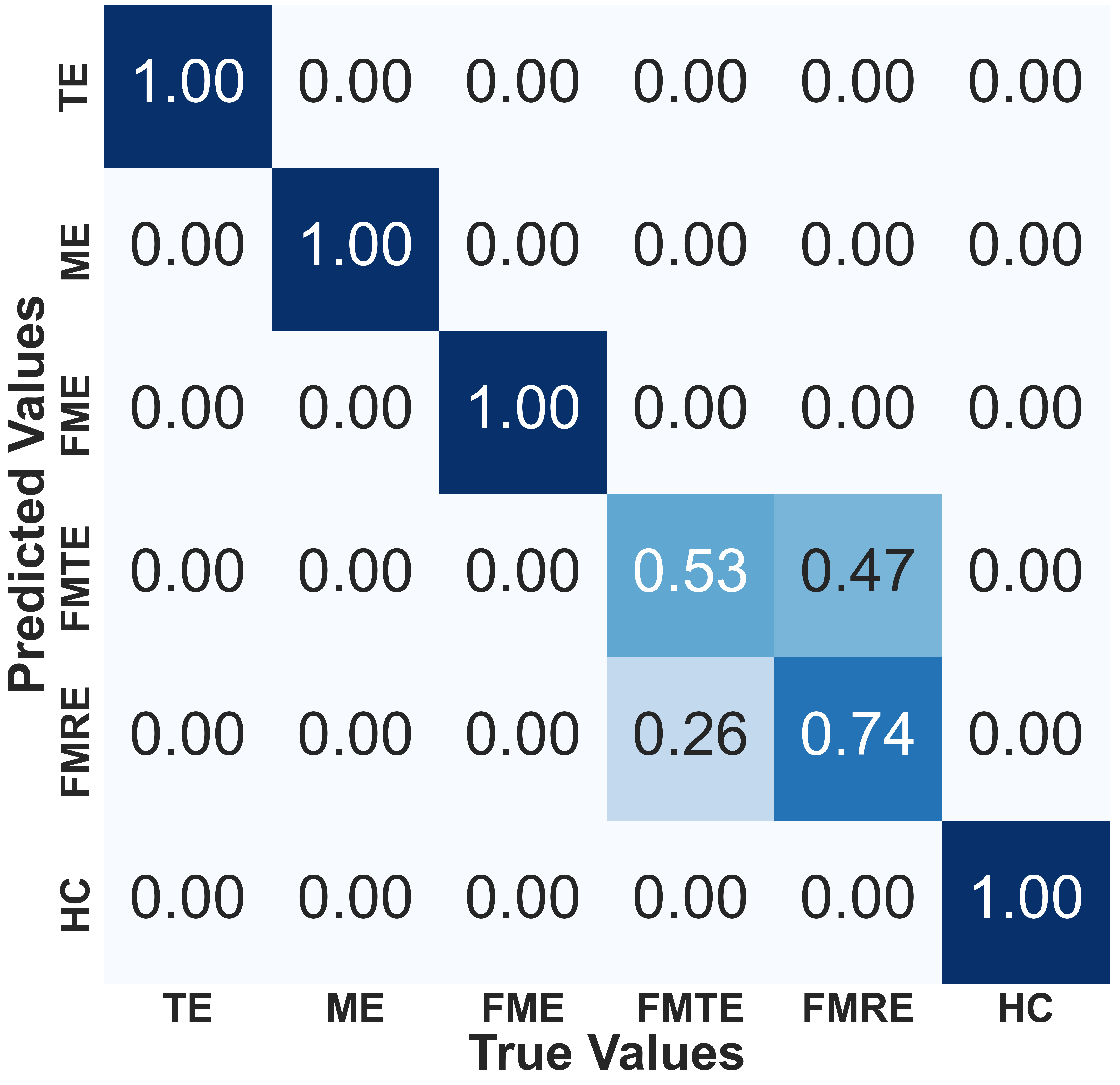}
		\caption{F1}\label{g1}
	\end{subfigure}	
	\begin{subfigure}{0.22\linewidth}
		\includegraphics[width=0.99\textwidth, height=0.9\linewidth]{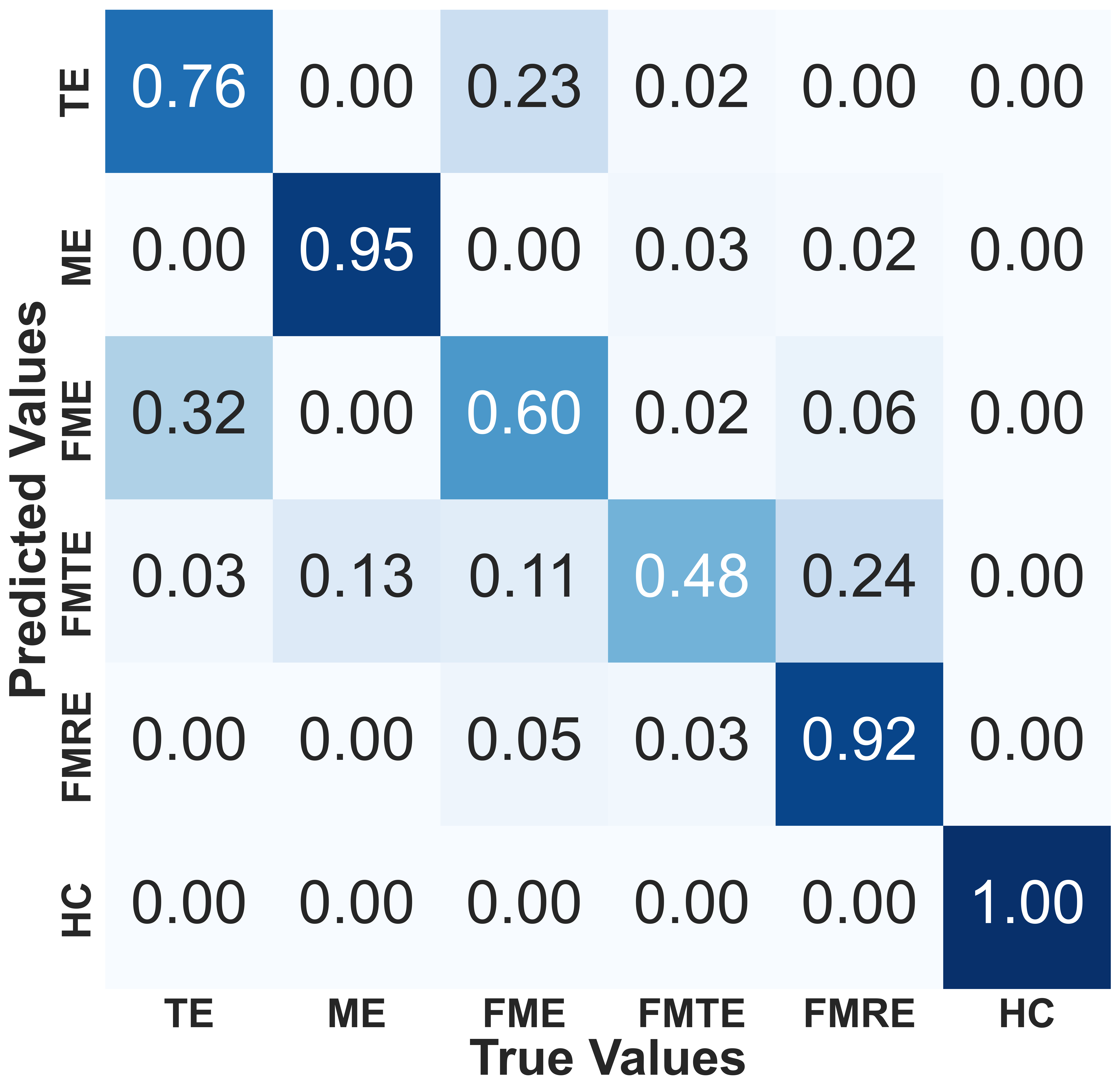}
		\caption{F2}\label{h1}
	\end{subfigure}	
	\caption{Confusion matrix of classifier ET and MEET} 
	\label{c1}
 \flushleft \footnotesize{TE = Thumb Extension, ME = Middle Extension, FME = Fore Middle Extension, FMTE = Fore Middle Thumb Extension, FMRE = Fore Middle Ring Extension, HC = Hand Close}
\end{figure*}

Consequently, our system outperforms the human gesture recognition method for all six movements in terms of diagonal percentage. Also, the off-diagonal element shows the gesture's misclassification amount.
\begin{figure}[]
\includegraphics[width=0.95\linewidth, height=0.42\textwidth]{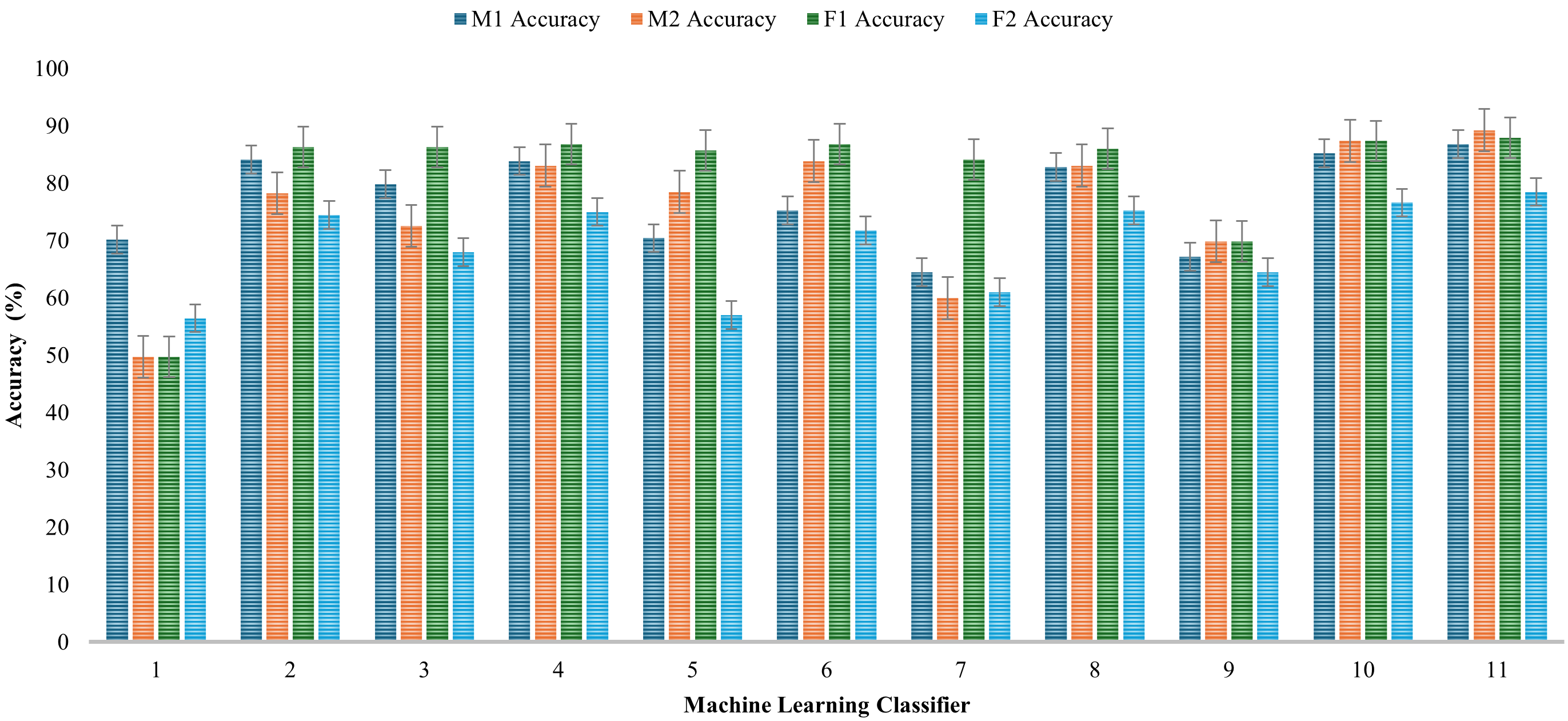}
\caption{Accuracy plot of all the classifiers with MEET}
\label{figh4}
\end{figure}
In Figure~\ref{c1}, a comparative analysis is presented between the ET classifier and the proposed MEET classifier's confusion matrix for all subjects. Figures~\ref{a1}-\ref{d1} illustrate the confusion matrices of subjects M1, M2, F1, and F2, respectively, using the second highest classifier, ET, in the study. Similarly, Figures~\ref{e1}-\ref{h1} display the highly accurate confusion matrices of the proposed MEET model for all subjects M1, M2, F1, and F2, respectively. In Figure~\ref{a1}, the first diagonal element value for the Hand gesture "TE" is 76\%, while for the same subjects using the MEET model (Figure 8e), there is an enhanced output with a diagonal value of 81\% for "TE," and similar improvements are observed for all the classes. 

\begin{wrapfigure}{r}{6cm}
\includegraphics[width=6cm]{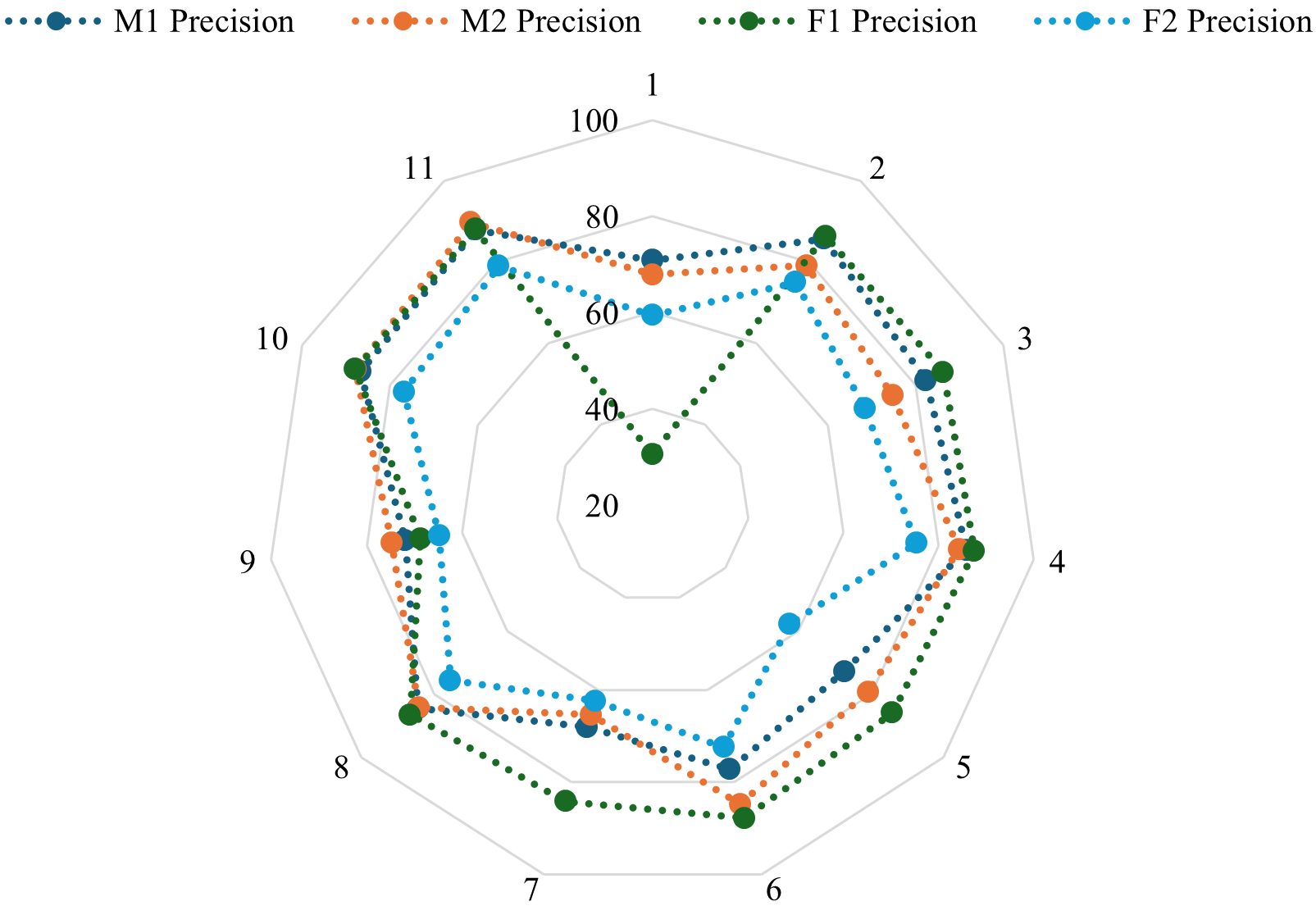}
\caption{Precision plot of all the classifiers with MEET}
\label{figh5}
\end{wrapfigure}
The remaining figures all demonstrate that the proposed architecture enhances the classification rate.
\begin{wrapfigure}{l}{6cm}
\includegraphics[width=6cm]{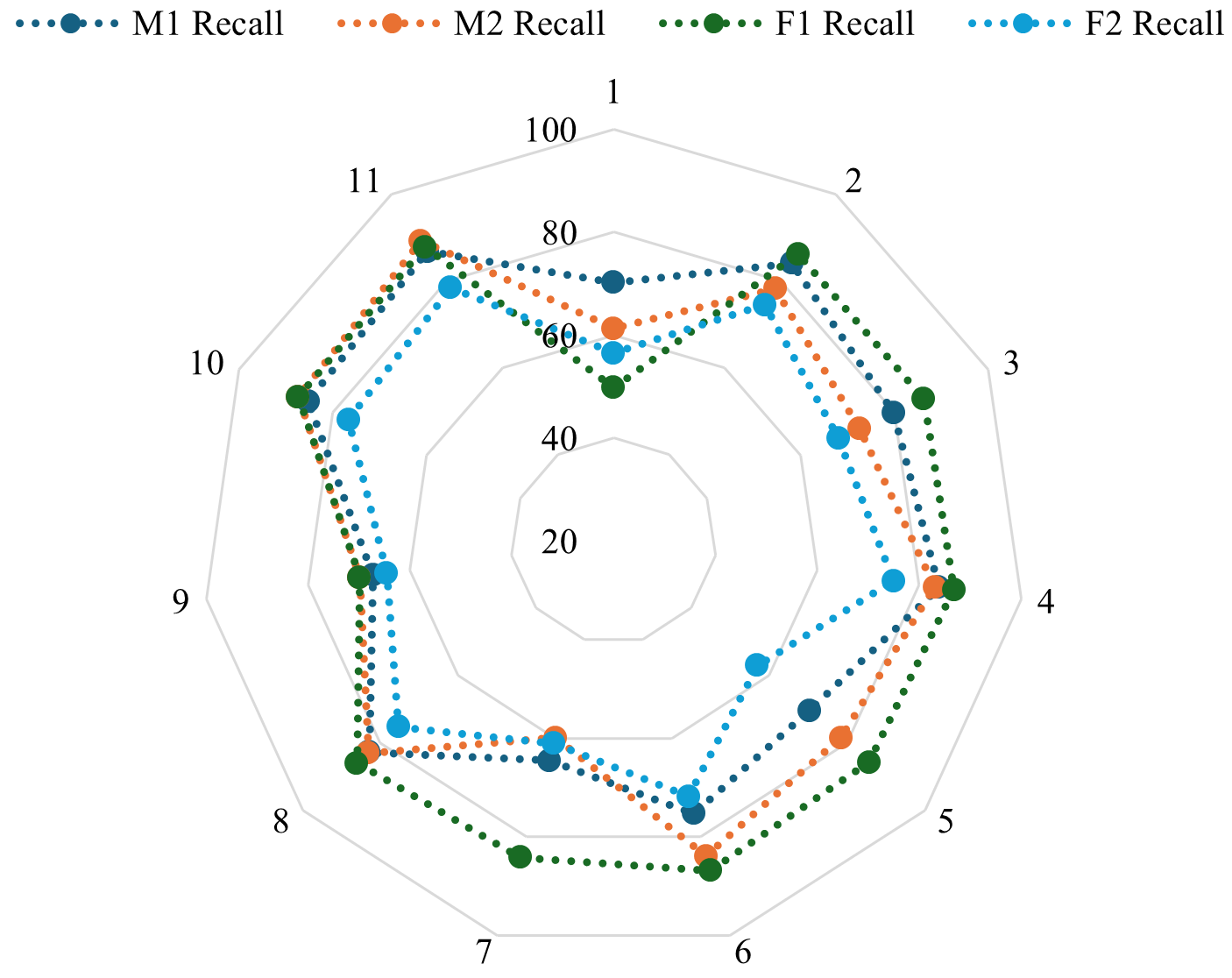}
\caption{Recall plot of all the classifiers with MEET}
\label{figh6}
\end{wrapfigure}

The table~\ref{t1} provides the $A_{cc}$ parameter for all of the models shown in Figure~\ref{figh4}. The graphic shows bar graphs, which may be used to visually display changes in the models using grouped or stacked bar charts. This allows for the comparison of data from various subjects using several models, which aids in the detection of trends and patterns. The decision-makers may readily understand critical information and make educated decisions based on insights gleaned from the chart. 

The bar plot graphically depicts the accuracy of all models used on the four subjects of the acquired data. Notably, the suggested model outperforms all other classifiers. The figure shows that the suggested model has an accuracy of 86.8\% for subject M1, 89.2\% for subject M2, 87.9\% for subject F1, and 78.4\% for subject F2.

\begin{wrapfigure}{r}{6cm}
\centering
\includegraphics[width=6cm]{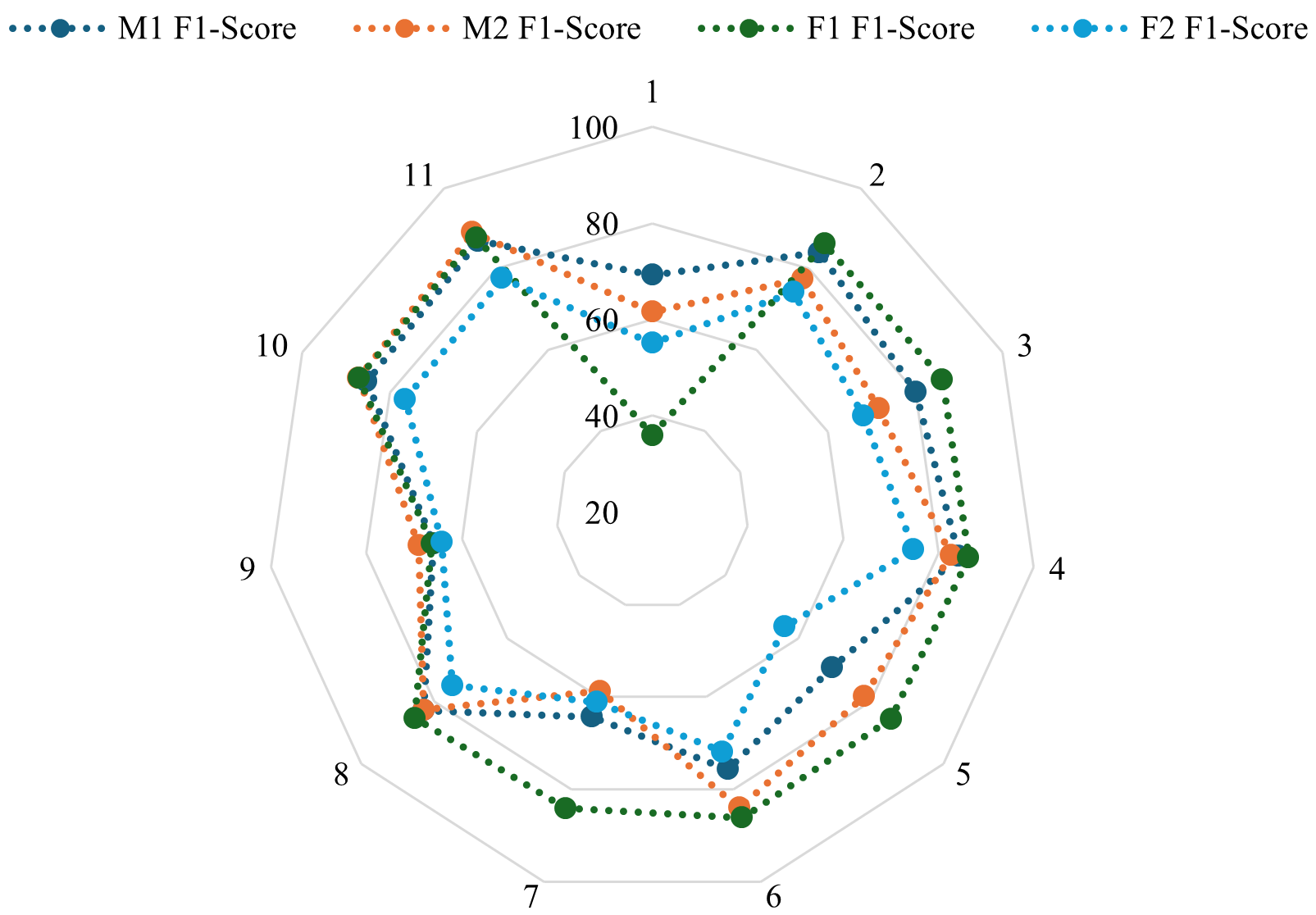}
\caption{F1-Score plot of all the classifiers with MEET}
\label{figh}
\end{wrapfigure}

The table's additional performance metrics include Precision, Recall, and F1-score, as shown in Figures~\ref{figh5},\ref{figh6}, and \ref{figh}. These findings are displayed as radar graphs, which depict multidimensional data representations and demonstrate the model's strengths and limitations across many areas. This graph enables the comparison of many topics with regard to a certain machine learning model.

Figure~\ref{figh5} examines the accuracy of the suggested model, which is an important metric in classification tasks. Precision measures the fraction of accurately predicted positive cases among all instances predicted as positive by the model.

This statistic is especially essential when the cost of false positives is high since it directly influences the dependability and trustworthiness of the models' predictions. According to the parameter analysis, the suggested model's accuracy for subjects M1, M2, F1, and F2 is 78\%, 79\%, 47\%, and 49\%, respectively. These percentages reflect the model's ability to identify relevant instances within each subject group reliably.

Similarly, when it comes to recollection across all subjects, the highly accurate model surpasses other models. Subjects M1, M2, F1, and F2 had recall values of 74\%, 78\%, 79\%, and 84\%, illustrating the model's exceptional ability to identify and retrieve relevant occurrences within each topic area reliably. This evaluation of recall measures demonstrates the model's greater performance in collecting all relevant events, reducing the likelihood of false negatives. By achieving high recall scores across several subjects, the model demonstrates its ability to identify significant information properly, enhancing its utility and reliability in a wide range of real-world scenarios.

Furthermore, the F1-score indicates the harmonic mean of accuracy and recall, allowing for a full evaluation of the model's performance across subjects M1, M2, F1, and F2. These participants had F1-scores of 45\%, 54\%, 56\%, and 78\%, respectively. This statistic combines accuracy and recall, providing a fair assessment of the model's ability to categorize cases while successfully reducing both false positives and false negatives. By attaining high F1-scores, the model displays its capacity to provide strong and consistent performance, establishing itself as a useful asset in a variety of practical applications and decision-making processes.

\section{Conclusion and Future Works}
This study proposed a new Mixture of Experts Extra Tree (MEET) as a machine learning approach that combines the strengths of Extra Trees. The advantages of MEET include the ability to improve classification accuracy by utilizing multiple Extra trees; each specialized in a particular aspect of the problem space. This study identified hand gesture movement using MEET and ten other ML classifiers. Four subjects two adult males and two adult females, were considered to acquire the hand gesture movement data sets. Seventeen significant features were extracted using time and frequency domain analysis and fed into ML classifiers. Results signify that the MEET classifier outperformed other classifiers and helped in achieving improved accuracy. Performance metrics such as accuracy, Precision, Recall, and F1-score for all four subjects indicated MEET classifier as an effective classifier for hand gesture movement identification. Future works in MEET could include:

\begin{enumerate}
    \item Investigating the use of MEET in other domains, such as natural language processing, image recognition, and time series analysis.
       \item Investigating the use of MEET in conjunction with other machine learning techniques, such as deep learning, to create hybrid models.
    \item Studying the impact of MEET on with the optimization of machine learning models with the hyperparameter tuning.
    \item Studying the impact of MEET on the overlapped data for the data split of individual expert models of machine learning.
\end{enumerate}

\bibliographystyle{ieeetr}
\bibliography{ref}

\clearpage

\end{document}